\documentclass[final,5p,twocolumn]{elsarticle}

\usepackage{hyperref}
\usepackage{lipsum}
\usepackage{amsmath}
\usepackage{amssymb}
\usepackage{flexisym}
\usepackage{xcolor}
\usepackage{float}

\usepackage{graphicx}
\usepackage{caption}
\usepackage{subcaption}

\usepackage{algorithm}
\usepackage{algorithmic}

\usepackage{balance}
\biboptions{sort&compress}
\hypersetup{draft}
\usepackage{wrapfig}
\usepackage{afterpage}

\makeatletter
\def\ps@pprintTitle{%
	\let\@oddhead\@empty
	\let\@evenhead\@empty
	\let\@oddfoot\@empty
	\let\@evenfoot\@oddfoot
}
\makeatother

\begin{document}
	
	\newcommand{\argmin}[1]{\underset{#1}{\operatorname{arg}\,\operatorname{min}}\;}
	\newcommand{\minmax}[1]{\underset{#1}{\operatorname{min}\,\operatorname{max}}\;}	
	\newcommand{\prob}[1]{p_#1} 	
	
	\newtheorem{example}{\bf Example}
	\newtheorem{definition}{\bf Definition}
	\newtheorem{theorem}{\bf Theorem}
	\newtheorem{claim}{\bf Claim}
	\newtheorem{proposition}{\bf Proposition}
	\newtheorem{lemma}{\bf Lemma}
	\newtheorem{remark}{\bf Remark}
	\newtheorem{collolary}{\bf Collolary}

	\newcommand{\Name}{AutoGAN-based Dimension Reduction for Privacy Preservation}
	\newcommand{\shortName}{AGDRPP}

	\begin{frontmatter}
		
		\title{ AutoGAN-based Dimension Reduction for Privacy Preservation}
		
		\author[add1]{Hung Nguyen}
		\ead{nsh@mail.usf.edu}
		
		\author[add1]{Di Zhuang}
		\ead{zhuangdi1990@gmail.com}
		
		\author[add2]{Peiyuan Wu}
		\ead{peiyuanwu@ntu.edu.tw}
		
		\author[add1]{Morris Chang}
		\ead{morrisjchang@gmail.com}
		
		\address[add1]{University of South Florida, USA}
		\address[add2]{National Taiwan University, Taiwan}

		\begin{abstract}
			\label{sec:abstract}

Protecting sensitive information against data exploiting attacks is an emerging research area in data mining. Over the past, several different methods have been introduced to protect individual privacy from such attacks while maximizing data-utility of the application. However, these existing techniques are not sufficient to effectively protect data owner privacy, especially in the scenarios that utilize visualizable data (e.g. images, videos) or the applications that require heavy computations for implementation. To address these problems, we propose a new dimension reduction-based method for privacy preservation. Our method generates dimension-reduced data for performing machine learning tasks and prevents a strong adversary from reconstructing the original data. We first introduce a theoretical approach to evaluate dimension reduction-based privacy preserving mechanisms, then propose a non-linear dimension reduction framework motivated by state-of-the-art neural network structures for privacy preservation. We conducted experiments over three different face image datasets (AT\&T, YaleB, and CelebA), and the results show that when the number of dimensions is reduced to seven, we can achieve the accuracies of 79\%, 80\%, and 73\% respectively and the reconstructed images are not recognizable to naked human eyes.			
		\end{abstract}
		
		\begin{keyword}
			Generative Adversarial Nets, Auto-encoder, neural-network, privacy preservation, dimension reduction, access control.
		\end{keyword}
		
	\end{frontmatter}	
	
	\section{Introduction}
	\label{sec:introduction}

Machine Learning (ML) is an important aspect of modern applications that rely on big data analytics (e.g., an on-line system collecting data from multiple data owners). However, these applications are progressively raising many different privacy issues as they collect different types of data on a daily basis. For example, many types of data are being collected in smart cities such as patient records, salary information, biological characteristics, Internet access history, personal images and so on. These types of data then can be widely used in daily recommendation systems, business data analysis, or disease prediction systems which in turn affect the privacy of individuals who contributed their sensitive data. Considering a multi-level access control system of a company using biometric recognition (e.g., face recognition, fingerprint) for granting permission to access data resources, the company staff members may concern their biological information being vulnerable to adversaries. Even though the utility of these biometric features can be effectively used in machine learning tasks for authentication purpose, leaking this information might lead to privacy breaches. For example, an adversary could utilize them to determine the members' identities.

Several tools and methods have been developed to preserve the privacy in machine learning applications, such as homomorphic encryption \cite{Bost2015,Emekci2007,Hesamifard2017}, secure multi-party computing \cite{Yao1986,Shamir1979}, differential privacy (DP) \cite{Chaudhuri,Zhang2012,Phan2016,Abadi2016,XiaoqianJiangZhanglongJiShuangWangNomanMohammedSamuelCheng2013}, compressive privacy \cite{zhuang2017fripal,GAP,DBLP:journals/corr/abs-1809-08911,DifferentialPrivacywithCompression,7815484, Kung, XieKun} and so on. Typically, differential privacy-based methods aim at preventing leaking individual information caused by queries. However, they are not designed to serve large number of queries since they require adding huge amount of noise to preserve privacy, thus significantly decreasing the ability to learn meaningful information from data. On the other hand, homomorphic encryption-based methods can be used to privately evaluate a function over encrypted data by a third party without accessing to plain-text data, hence the privacy of data owners can be protected. However, due to the high computational cost and time consumption, they may not work with a very large dataset, normally required in ML applications.

In this study, we consider an access control system collecting dimension-reduced face images of staff members to perform authentication task and to provide permission for members who would like to access company's data resources (Figure \ref{fig:attackmodel}). We propose a non-linear dimension reduction framework to decrease data dimension for the authentication purpose mentioned above and to protect against an adversary from reconstructing member images. Firstly, we introduce $\epsilon$-DR Privacy as a theoretical tool for dimension reduction privacy evaluation. It evaluates the reconstruction distance between original data and reconstructed data of a dimension reduction (DR) mechanism. This approach encourages a DR mechanism to enlarge the distance as high distance yields high level of privacy. While other methods such as differential privacy-based methods rely on inference uncertainty to protect sensitive data, $\epsilon$-DR Privacy is built on reconstruction error to evaluate privacy. Therefore, unlike differential privacy methods,  $\epsilon$-DR Privacy is not negatively impacted by the number of queries. Secondly, as detailed in Section \ref{methodology}, we recommend a privacy-preserving framework Autoencoder Generative Adversarial Nets-based Dimension Reduction Privacy (AutoGAN-DRP) for enhancing data owner privacy and preserving data utility. The \textit{utility} herein is evaluated via machine learning task performance (e.g., classification accuracy).

Our dimension reduction (DR) framework can be applied to different types of data and used in several practical applications without heavy computation of encryption and impact of query number. The proposed framework can be applied directly to the access control system mentioned above. More elaboratively, face images are locally collected, nonlinearly compressed to achieve DR, and sent to the authentication center. The server then performs classification tasks on the dimension-reduced data. We assume the authentication server is semi-honest, that is to say it does not deviate from authenticating protocols while being curious about a specific member's identity. Our DR framework is designed to resist against reconstruction attacks from a strong adversary who obtains the training dataset and the transformation model. 

During the stage of experiments, we implemented our framework to evaluate dimension-reduced data in terms of accuracy of the classification tasks, and we attempted to reconstruct original images to examine the capacity of adversaries. We performed several experiments on three facial image datasets in both gray-scale and color, i.e., \textit{the Extended Yale Face Database B} \cite{GeBeKr01}, \textit{AT}\&\textit{T} \cite{341300}, and \textit{CelebFaces Attributes Dataset (CelebA)} \cite{celeba}. The experiment results illustrate that with only seven reduced dimensions our method can achieve accuracies of 93\%, 90\%, and 80\% for AT\&T, YaleB, and CelebA respectively. Further, our experiments show that at the accuracies of 79\%, 80\% and 73\% respectively, the reconstructed images could not be recognized by human eyes. In addition, the comparisons shown in Section \ref{AutoGAN_DP_PCA} also illustrate that AutoGAN-DRP is more resilient to reconstruction attacks compared to related works.
\\\\
\textbf{Our work has two main contributions:}
\begin{itemize} 
\item To analytically support privacy guarantee, we introduce $\epsilon$-DR Privacy as a theoretical approach to evaluate privacy preserving mechanism.
\item We propose a non-linear dimension reduction framework for privacy preservation motivated by Generative Adversarial Nets \cite{Goodfellow2014} and Auto-encoder Nets \cite{Baldi2012}. 
  
\end{itemize}
The rest of our paper is organized as follows. Section 2 summarizes state-of-the-art privacy preservation machine learning (PPML) techniques and reviews knowledge of deep learning methods including generative adversarial neural nets and Auto-encoder. Section 3 describes the privacy problem through a scenario of a facial recognition access control system, introduces the definition of $\epsilon$-DR Privacy to evaluate DR-based privacy preserving mechanisms, and presents our framework \Name. Section 4 presents and discusses our experiment results over three different face image datasets. Section 5 compares AutoGAN-DRP to a similar work GAP in terms of reconstruction error and classification accuracy. Section 6 demonstrates reconstructed images over AutoGAN-DRP and other privacy preservation techniques (i.e., Differential Privacy and Principle Component Analysis). Finally, the conclusion and future work are mentioned in Section 7.

	\section{Related Work}
	
	\subsection{Literature Review}
	
\textbf{Cryptographic approach:} This approach usually applies to the scenarios where the data owners do not wish to expose their plain-text sensitive data while asking for machine learning services from a third-party. The most common tool used in this approach is fully homomorphic encryption that supports multiplication and addition operations over encrypted data, which enabling the ability to perform a more complex function. However, the high cost of the multiplicative homomorphic operations renders it difficult to be applied on machine learning tasks. In order to avoid multiplicative homomorphic operations, additive homomorphic encryption schemes are more widely used in privacy preserving machine learning (PPML). However, the limitation of the computational capacity in additive homomorphic schemes narrows the ability to apply on particular ML techniques. Thus, such additive homomorphic encryption-based methods in \cite{Bost2015,Emekci2007,Raphael,Hesamifard}  are only applicable to simple machine learning algorithms such as decision tree and naive bayes. In Hesamifard's work \cite{Hesamifard2017},the fully homomorphic encryption is applied to perform deep neural networks over encrypted data, where the non-linear activation functions are approximated by polynomials. 
  
In secure multi-party computing (SMC), multiple parties collaborate to compute functions without revealing plain-text to other parties. A widely-used tool in SMC is garbled circuit \cite{Yao1986}, a cryptographic protocol carefully designed for two-party computation, in which they can jointly evaluate a function over their sensitive data without the trust of each other. In \cite{Al-rubaie}, Mohammad introduced a SMC protocol for principle component analysis (PCA) which is a hybrid system utilizing additive homomorphic and garbled circuit. In secret sharing techniques \cite{Shamir1979}, a secret \textbf{s} is distributed over multiple pieces \textbf{n} also called \textit{shares}, where the secret can only be recovered by a sufficient amount of \textbf{t} \textit{shares}. A good review of secret sharing-based techniques and encryption-based techniques for PPML is given in \cite{Pedersen2007}. Although these encryption-based techniques can protect the privacy in particular scenarios, their computational cost is a significant concern. Furthermore, as \cite{Pedersen2007} elaborated, the high communication cost also poses a big concern for both techniques.
      
\textbf{Non-Cryptographic approach:} 
Differential Privacy (DP) \cite{Dwork2006} aims to prevent membership inference attacks. DP considers a scenario that an adversary infers a member's information based on the difference of outputs of a ML mechanism before and after the member join a database. The database with the member's information and without the member's information can be considered as two neighbor databases which differ by at most one element. DP adds noise to the outputs of the ML mechanism to result in similar outputs from the two neighbor databases. Thus, adversaries cannot differentiate the difference between the two databases. A mechanism M satisfies $\epsilon$-differential privacy if for any two neighbor databases $D$ and $D'$, and any subset S of the output space of M satisfies 
$ Pr[M(D) \in S] \leq e^{\epsilon}  Pr[M(D') \in S]  $. 
The similarity of query outputs protects a member information from such membership inference attacks. The \textit{similarity} is guaranteed by the parameter $\epsilon$ in a mechanism in which the smaller $\epsilon$ provides a better level of privacy preservation. \cite{Chaudhuri, Zhang2012,NIPS2008, Wu, MYang} propose methods to guarantee $\epsilon$-differential privacy by adding noise to outcome of the weights $w^*= w + \eta$, where $\eta$ drawn from Laplacian distribution and adding noise to the objective function of logistic regression or linear regression models. \cite{Phan2016,Abadi2016} satisfy differential privacy by adding noise to the objective function while training a deep neural network using stochastic gradient descent as the optimization algorithm. 

In addition, there are existing works proposing differential privacy dimension reduction. One can guarantee $\epsilon$-differential privacy by perturbing dimension reduction outcome. Principal component analysis (PCA) whose output is a set of eigenvectors is a popular method in dimension reduction. The original data is then represented by its projection on those eigenvectors, which keeps the largest variance of the data. One can reduce the data dimension by eliminating insignificant eigenvectors which contain less variance, and apply noise on the outcome to achieve differential privacy\cite{XiaoqianJiangZhanglongJiShuangWangNomanMohammedSamuelCheng2013}. However, the downside of these methods is that they are designed for specific mechanisms and datasets and not working well with the others. For example, record-level differential privacy is not effectively used with image dataset as shown in \cite{Hitaj2017}. Also, the amount of added noise is accumulative based on the number of queries so that this approach usually leads to low accuracy results with a high number of queries. 

Similar to our work, Generative Adversarial Privacy (GAP) \cite{GAP} is a perturbation method utilizing the minimax algorithm of Generative Adversarial Nets to preserve privacy and to keep utility of image datasets. GAP perturbs data within a specific $l_2$ distance constraint between original and perturbed data to distort private class labels and at the same time preserve non-private class labels. However, it does not protect the images themselves, and an adversary can visually infer private label (e.g., identity) from images. In contrast, our method protects an image by compressing it into a few dimension vector and then transferring without clearly exposing the original image.

	\subsection{Preliminaries}
	To enhance the distance between original and reconstructed data in our DR system, we utilize the structure of Generative Adversarial Network (GAN) \cite{Goodfellow2014} for data perturbation and deep Auto-encoder \cite{Baldi2012} for data reconstruction. The following sections briefly review Auto-encoder and GAN.
	\subsubsection{Auto-encoder}
	Auto-encoder is aimed at learning lower dimension representations of unsupervised data. Auto-encoder can be used for denoising and reducing data dimension. It can be implemented by two neural network components: \textit{encoder} and \textit{decoder}. The \textit{encoder} and \textit{decoder} perform reverse operations. The input of the \textit{encoder} is the original data while the output of the \textit{decoder} is expected to be similar to the input data. The middle layer extracts latent representation of original data that could be used for dimension reduction. An Auto-encoder training process can be described as a minimization problem of the auto-encoder's loss function $\mathcal{L(\cdot)}$:
	\begin{equation}
	\mathcal{L}(x,g(f(x)))
	\end{equation}
	where x is input data, f($\cdot$) is an encoding function, and g($\cdot$) is a decoding function.
	
	\subsubsection{GAN}
	Generative Adversarial Nets is aimed at approximating distribution $p_d$ of a dataset via a generative model. GAN simultaneously trains two components \textit{generator} $G$ and \textit{discriminator} $D$, and the input of $G$ is sampled from a prior distribution $p_z(z)$ through which $G$ generates fake samples similar to the real samples. At the same time, $D$ is trained to differentiate between fake samples and real samples, and send feedback to $G$ for improvement. GAN can be formed as a two-player minimax game with value function V(G,D):
	
	\begin{equation}
	\begin{split}
	\min_G \max_D{V(G,D)} = & E_{x \sim \prob{d}} [log(D(x))] + \\& E_{z\sim \prob{z} }[log( 1 - D( G(z) ))]
	\end{split}
	\end{equation}
	
	The two components, \textit{Generator} and \textit{Discriminator} can be built from neural networks (e.g., fully connected neural network, convolutional neural network). The goal of G is to reduce the accuracy of D. Meanwhile, the goal of D is to differentiate fake samples from real samples. These two components are trained until the discriminator cannot distinguish between generated samples and real samples.

	\section{Methodology} \label{methodology}
	
	In this section, we first describe the problem and threat model, then we introduce a definition of DR-Privacy and our dimensionality reduction method (AutoGAN-DRP).

\begin{figure}[h]
	\includegraphics[width=\linewidth]{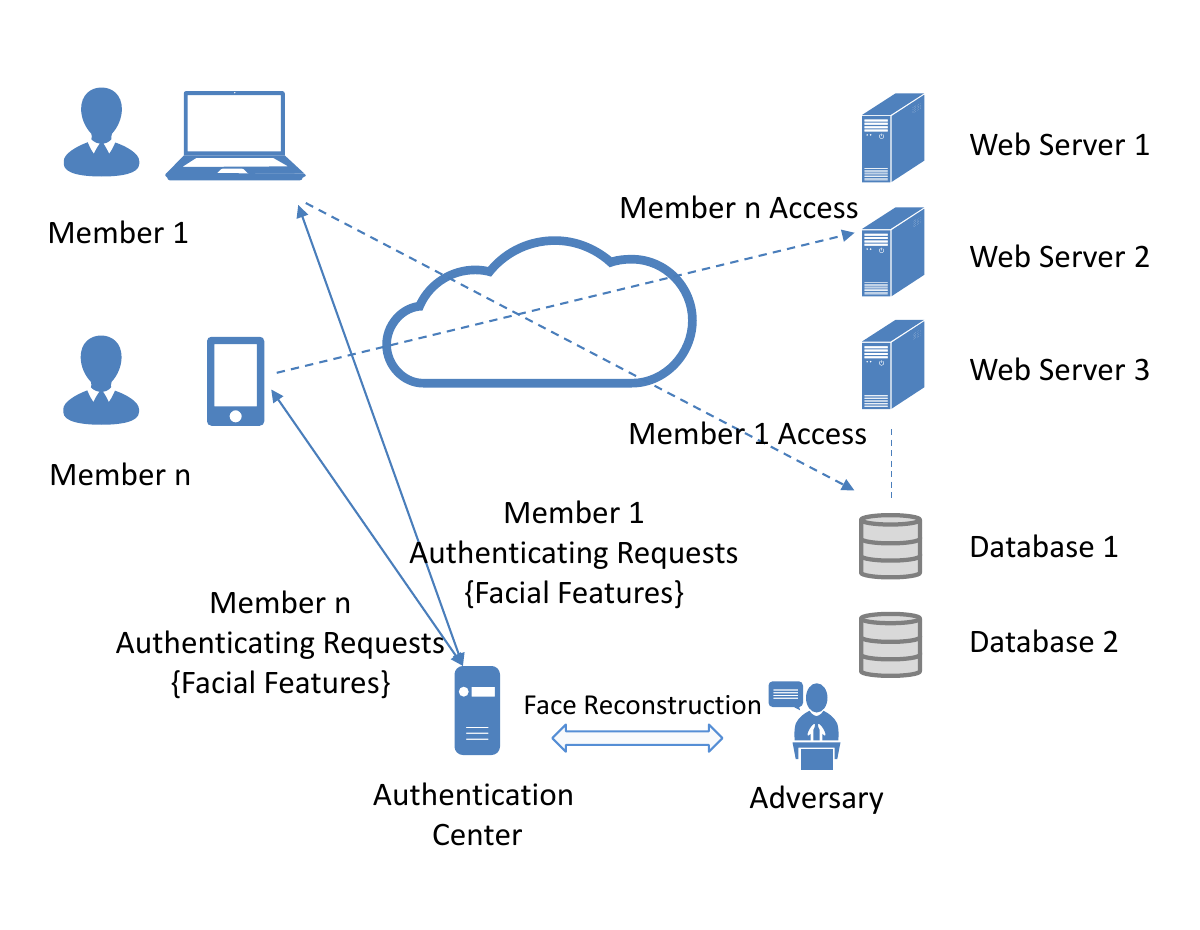}
	\captionsetup{justification=centering}
	\caption{Attack Model}
	\label{fig:attackmodel}
\end{figure} 

\subsection{Problem statement}
We introduce the problem through the practical scenario mentioned in Section \ref{sec:introduction}. Figure \ref{fig:attackmodel} briefly describes the entire system in which staff members (clients) in a company request access to company resources, such as websites and data servers through a face recognition access control system. For example, if member n requests to access web server 2, the local device first takes a facial photo of the member by an attached camera, locally transforms it into lower dimension data, and sends to an authentication center. The authentication server then obtains the low dimensional data and determines member access eligibility by using a classifier without clear face images of the requesting member. We consider that the system has three levels of privileges (i.e., single level, four-level, eight-level) corresponding to three groups of members. We assume the authentication server is semi-honest (it obeys work procedure but might be used to infer personal information). If the server is compromised, an adversary in the authentication center can reconstruct the face features to achieve plain-text face images and determine members' identity.  
\subsection{Threat Model}
In the above scenario, we consider that a strong adversary who has access to the model and training dataset attempts to reconstruct the original face images for inferring a specific member's identity. Our attack model can be represented in Figure \ref{fig:attackmodel}. The adversary utilizes training data and facial features to identify a member identity by reconstructing the original face images using a reconstructor in an auto-encoder. Rather than using fully connected neural network, we implement the auto-encoder by convolutional neural network which more effective for image datasets. Our goal is to design a data dimension reduction method for reducing data dimension and resisting full reconstruction of original data.

	\subsection{$\epsilon$-Dimension Reduction Privacy ($\epsilon$-DR Privacy)} \label{theory}

 \begin{figure}[H]
	\centering
	\includegraphics[width=0.4\textwidth]{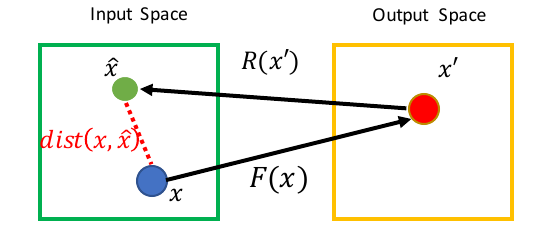}
	\caption{DR projection and reconstruction.} \label{fig:DRP}
\end{figure} 

We introduce the Dimension Reduction Privacy (DR-Privacy), and define a formal definition of the $\epsilon$-DR Privacy to mathematically quantify/evaluate the mechanisms designed to preserve the DR-Privacy 
via dimension reduction. The DR-Privacy aims to achieve privacy-preserving via dimension reduction, which refers to transforming the data into a lower dimensional subspace, such that the private information is concealed while the underlying probabilistic characteristics are preserved, which can be utilized for machine learning purposes. To quantify the DR-Privacy and guide us to design such DR functions, we define $\epsilon$-DR Privacy as follows.

{\bf Definition 1: ($\epsilon$-DR Privacy)} A Dimension Reduction Function $F(\cdot)$ satisfies $\epsilon$-DR Privacy if for each i.i.d. $m$-dimension input sample $x$ drawn from the same distribution $D$, and for a certain distance measure $dist(\cdot)$, we have
\begin{equation}
\begin{split} \label{eq:modularity}
\mathbb{E}[dist(x, \hat{x})] \geq \epsilon
\end{split}
\end{equation}
where $\mathbb{E}[\cdot]$ is the expectation, $\epsilon \geq 0$, $x^{\prime}=F(x)$, $\hat{x}=R(x^{\prime})$, and $R(\cdot)$ is the Reconstruction Function.

For instance, as shown in Fig.~\ref{fig:DRP}, given original data $x$, our framework utilizes certain dimension reduction function $F(x)$ to transform the original data $x$ into the transformed data $x^{\prime}$. The adversaries aim to design a corresponding reconstruction function $R(x^{\prime})$ such that the reconstructed data $\hat{x}$ would be closed/similar to the original data $x$. DR-Privacy aims to design/develop such dimension reduction functions, that the distance between the original data and its reconstructed data would be large enough to protect the privacy of the data owner.

\begin{figure}
	\includegraphics[width=\linewidth, trim= 8.5cm 5.5cm 8.5cm 5.5cm, clip=true]{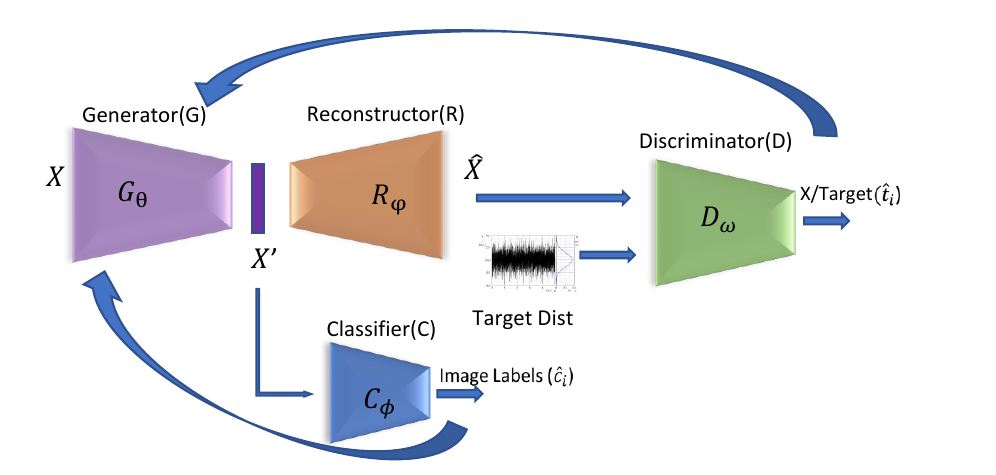}
	\captionsetup{justification=centering}
	\caption{AutoGAN-DRP}
	\label{fig:eGAN}
\end{figure}

\subsection{AutoGAN-based Dimension Reduction for Privacy Preserving (AutoGAN-DRP)}

We propose a deep learning framework for transforming face images to low dimensional data which is hard to be fully reconstructed. The framework can be presented in Figure \ref{fig:eGAN}. We leverage the structure of an auto-encoder \cite{Baldi2012} which contains encoder and decoder (in this work, we called them generator and re-constructor) in order to reduce data dimension. More specifically, the low dimensional representations are extracted from the middle layer of the auto-encoder (the output of the generator). The dimension-reduced data can be sent to the authentication server as an authentication request. We consider an adversary as a re-constructor implemented by a decoder. To resist against fully reconstructing images, the framework utilizes a discriminator in GAN \cite{Goodfellow2014} to direct reconstructed data to a designated target distribution with an assumption that the target distribution is different from our data distribution. In this work, the target distribution is sampled from Gaussian distribution and the mean is the average of training data. After projecting data into a lower dimension domain, the re-constructor is only able to partially reconstruct the data. Therefore, the adversary might not be able to recognize an individual's identity. To maintain data utility, we also use feedback from a classifier. The entire framework is designed to enlarge the distance between original data and its reconstruction to preserve individual privacy and retain significant data information. The dimension-reduced transformation model is extracted from the framework and provided to clients for reducing their face image dimensions. The classification model will be used in an authentication center that classifies whether a member's request is valid to have access (1) or not (0). 
 
We formulate the problem as follows:
Let $X$ be the public training dataset. $(x_i, y_i)$ is the $i$th sample in the dataset in which each sample $x_i$ has $d$ features and a ground truth label $y_i$. The system is aimed at learning a dimension reduction transformation $F(\cdot)$ which transforms the data from $d$ dimensions to $d'$ dimensions in which $d' \ll d $. Let $X'$ be the dataset in lower dimension domain. The dimension-reduced data should keep significant information to work with different types of machine learning tasks and should resist against the reconstruction or inference from data owner information.   

Our proposed framework is designed to learn a DR function $F(\cdot)$ that projects data onto low dimension space and preserves privacy at certain value of $\epsilon$. The larger distance implies higher level of privacy. Figure \ref{fig:eGAN} presents our learning system in which the dimension-reduced data $X'$ is given by a generator $G$. Since $X'$ is expected to be accurately classified by a classifier $C$, the generator improves by receiving feedback from the classifier via the classifier's loss function $\mathcal{L}_C$. We use a binary classifier for single-level authentication system and multi-class classifiers for multi-level authentication system. The classifier loss function is defined as the cross entropy loss of the ground truth label $y$ and predicted label $\hat y$ as follows.

\begin{equation}
\mathcal{L}_C=-\sum_{i=1}^n\sum_{j=1}^m y_{ij} \log(\hat y_{ij})  
\label{C_loss}
\end{equation}
where $m$ denotes the number of classes and $n$ denotes the number of samples.

To evaluate data reconstruction and enlarge the reconstruction distance, a re-constructor $R$ is trained as a decoder in an auto-encoder and sends feedback to the generator via its loss function $\mathcal{L}_R$. The re-constructor plays its role as an aggressive adversary attempting to reconstruct original data by using known data. The loss function of $R$ is the mean square error of original training data ($x$) and reconstructed data ($\hat x$), as displayed in (\ref{R_loss}).
\begin{align}
\mathcal{L}_R = \sum_{i=1}^n{(x_i - \hat x_i)^2} 
\label{R_loss}
\end{align}   
 
To direct the reconstructed data to a direction that reveals less visual information, the generator is trained with a discriminator $D$ as a minimax game in GAN. The motivation is to direct reconstructed data to a certain target distribution (e.g., normal distribution). To ensure a distance, the target distribution should be different to training data distribution. The discriminator aims to differentiate the reconstructed data from samples of the target distribution. The loss function of $D$ ($\mathcal{L}_D$) can be defined as a cross-entropy loss of ground truth labels (0 or 1) $t$ and prediction labels $\hat t$ shown in (\ref{D_loss}).

\begin{equation}
\mathcal{L}_D = -\sum_{i=1}^n{(t_i\log(\hat t_i) + (1 - t_i)\log(1 - \hat t_i))} 
\label{D_loss}
\end{equation}

The optimal generator parameter $\theta^*$ is given by the optimization problem of the generator loss function  $\mathcal{L}_G$:
\begin{equation}
\underset{\theta}{minimize} \; \mathcal{L}_G(\theta) = \alpha \min\limits_{\phi}{\mathcal{L}_C} - \beta\min\limits_{\omega}{\mathcal{L}_D}
 -\gamma\min\limits_{\varphi}{\mathcal{L}_R} + \mathcal{C}(\epsilon)
 \label{eqn:G_loss}
\end{equation}  
where $\theta$, $\phi$, $\omega$, and $\varphi$ are the model parameters of the generator, classifier, discriminator, and re-constructor respectively. $\alpha$, $\beta$, and $\gamma$ are weights of components in the objective function of the generator and can be freely tuned. $\mathcal{C}(\epsilon)$ is a constraint function with respect to hyper-parameter $\epsilon$, as to be elaborated in the following subsection.

\subsection{Optimization With Constraint}
In order to meet a certain level of reconstruction distance, we consider the constrained problem:

\begin{equation}
\begin{array}{l}
\; \; \; \;\underset{\theta}{minimize} \;\mathcal{L}_G(\theta) \\ 
s.t \; \; \;  \mathbb{E}_{x \sim p_d}[dist(x, \hat{x})] \leq \epsilon  
\end{array}
\label{constr}
\end{equation}

The optimization problem above can be approximated as an unconstrained problem \cite{pauljensen}:
\begin{equation} 
\underset{\theta}{minimize} \; ( \mathcal{L}_G(\theta) + \gamma \mathcal{C}(\epsilon) )  
\end{equation}
 where $\gamma$ is a penalty parameter and $\mathcal{C}$ is a penalty function 

\begin{equation} 
\mathcal{C}(\epsilon) = \max(0, \mathbb{E}_{x \sim p_d}[dist(x, \hat{x})] -\epsilon)
\end{equation}
Note that $\mathcal{C}$ is nonnegative, and $\mathcal{C}(\theta)=0$ iff the constraint in (\ref{constr}) is satisfied.

\subsection{Training Algorithms}

\begin{algorithm}[h]
	\caption{Algorithm for stochastic gradient descent training of $\epsilon$ -DR Privacy.}
	\begin{algorithmic}[1]
		\renewcommand{\algorithmicrequire}{\textbf{Input:}}
		\renewcommand{\algorithmicensure}{\textbf{Output:}}
		\REQUIRE Training dataset $X$. \\Parameter: learning rate $\alpha_r,\alpha_d,\alpha_c,\alpha_g $, 
		training steps $n_r,n_d,n_c,n_g$ \\
		A constraint for $\epsilon$-DR 
		\ENSURE  Transformation Model
		\\ \textit{Initialization.} 
		\FOR {$n$ global training iterations}
		\STATE  Randomly sample a mini batch from target distribution and label $\boldsymbol{t}$.\\
		\STATE  Randomly sample mini batch of data $\boldsymbol x $ and corresponding label $\boldsymbol{y}$  
		
		\FOR{$i = 0 $ to $n_r$ iterations}
		\STATE	Update the Reconstruction:\\
		$ \varphi_{i+1} = \varphi_{i} - \alpha_r \nabla_\varphi{\mathcal{L}_R(\varphi_{i} ,\boldsymbol{x} ) }	$\\
		\ENDFOR	
		\FOR{$j = 0 $ to $n_d$ iterations}
		\STATE	Update the Discriminator parameter:\\
		$ \omega_{j+1} = \omega_{j} - \alpha_d \nabla_\omega{\mathcal{L}_D(\omega_{j} ,\boldsymbol{x,t} ) }	$
		\ENDFOR	
		\FOR{$k = 0 $ to $n_c$ iterations}
		\STATE	Update the Classifier parameter:\\
		$ \phi_{k+1} = \phi_{k} - \alpha_c \nabla_\phi{\mathcal{L}_C(\phi_{k} ,\boldsymbol{x,y} ) }	$
		\ENDFOR	
		\FOR{$l = 0 $ to $n_g$ iterations}
		\STATE	Update the Generator parameter:\\
		$\theta_{l+1} = \theta_{l} - \alpha_g \nabla_\theta{\mathcal{L}_G(\theta_{l} ,\boldsymbol{x,t,y} ) }$
		\ENDFOR					
		\ENDFOR
		
		\RETURN 
	\end{algorithmic}
	\label{alg}
\end{algorithm}

Algorithm \ref{alg} describes the training process of AutoGAN-DRP. The framework contains four components, and they are trained one by one (lines 4-15) within one global training step. After sampling batches from target distribution and data for inputs of the models (lines 2-3), we  then train the four components. First, the re-constructor is trained in $n_r$ iterations while other components' parameters are fixed (lines 4-6). Second, the discriminator is trained (lines 7-9). Third, the classifier is trained in $n_c$ iterations (lines 10-12). Fourth, the generator is trained in $n_g$ iterations (lines 13-15). After training each component in their number of local training steps, the above training process is repeated until it reaches the number of global training iterations (lines 1-16). In our setting, the numbers of local training iterations ($n_c$, $n_r$, $n_d$, $n_g$ ) are much smaller than the number of global iterations $n$.

	\section{Experiments and Discussion}

\begin{table*}
	\centering
	\includegraphics[width=0.9\linewidth, trim=1cm 2.5cm 0.1cm 2.5cm, clip=true]{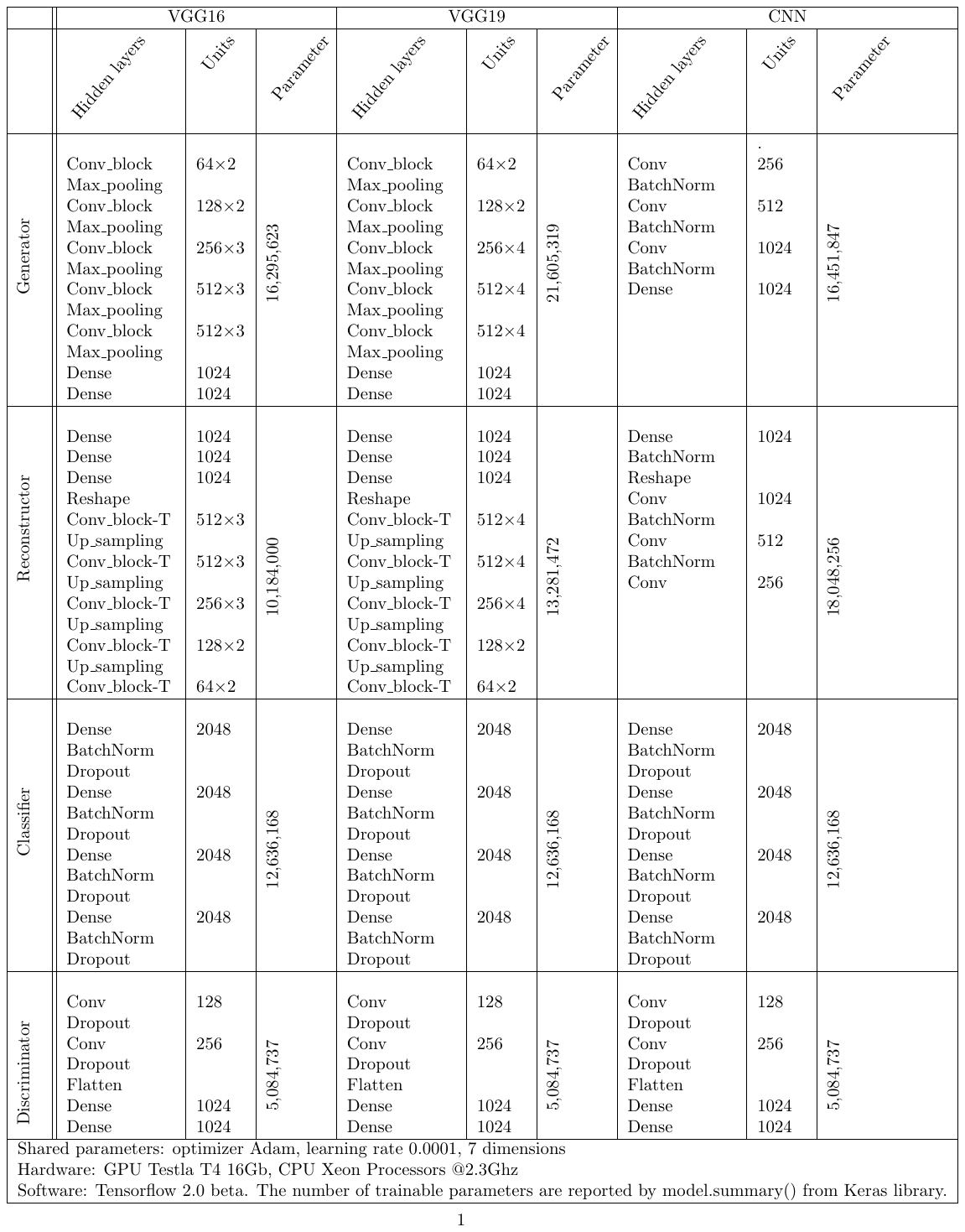}
	\caption{Implementation information}
	\label{table:implementation}
\end{table*}

\begin{figure*}[ht!]
	\begin{subfigure}{.33\textwidth}
		\includegraphics[width=\linewidth, trim=3.8cm 8cm 4cm 8cm, clip=true]{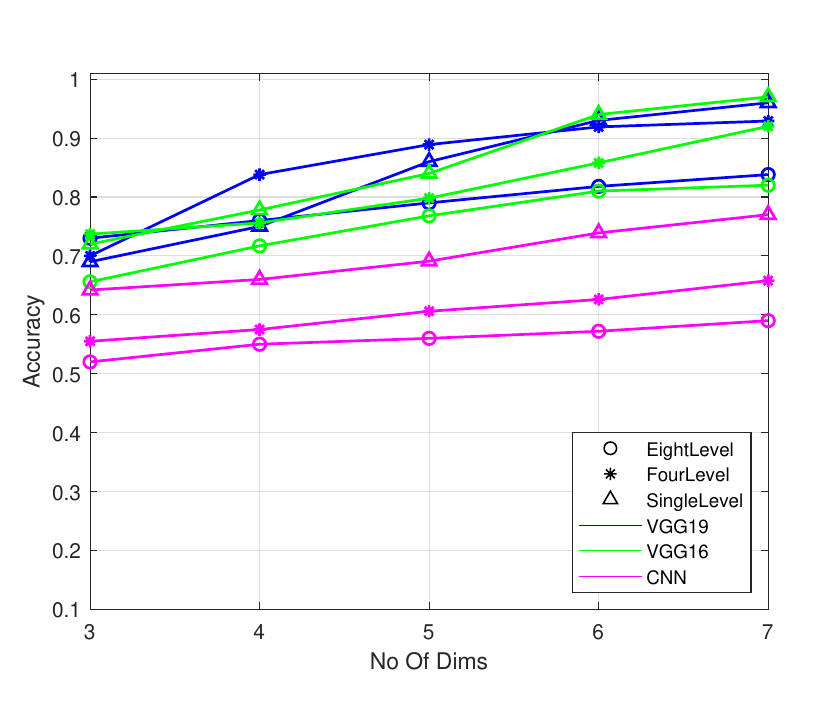}
		\captionsetup{justification=centering}
		\caption{ AT\&T}
		\label{fig:att_acc}
	\end{subfigure}
	\begin{subfigure}{.33\textwidth}
		\includegraphics[width=\linewidth, trim=3.8cm 8cm 4cm 8cm, clip=true]{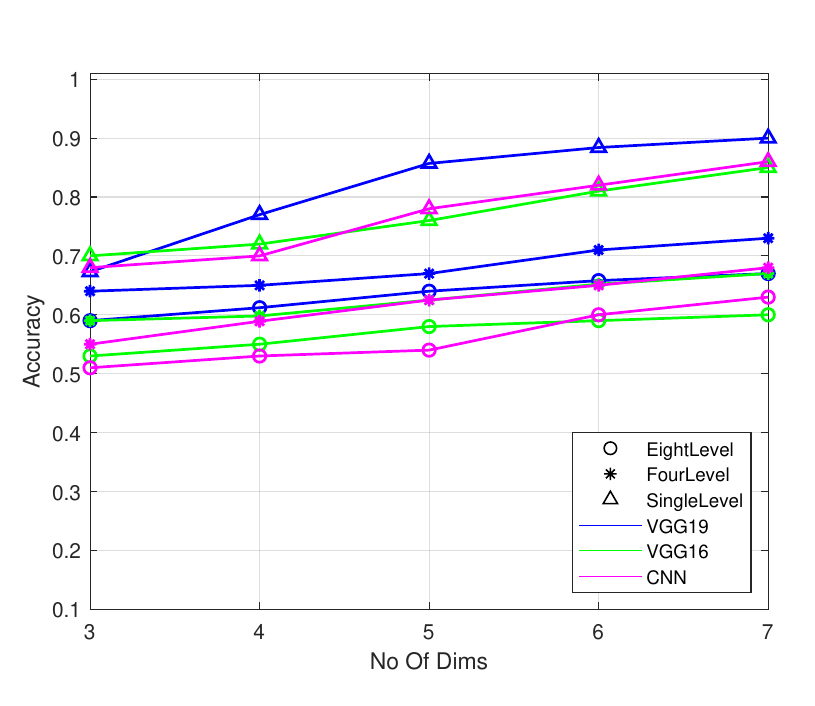}
		\captionsetup{justification=centering}
		\caption{Yale\_B}
		\label{fig:yale_acc}
	\end{subfigure}
	\begin{subfigure}{.33\textwidth}
		\includegraphics[width=\linewidth, trim=3.8cm 8cm 4cm 8cm, clip=true]{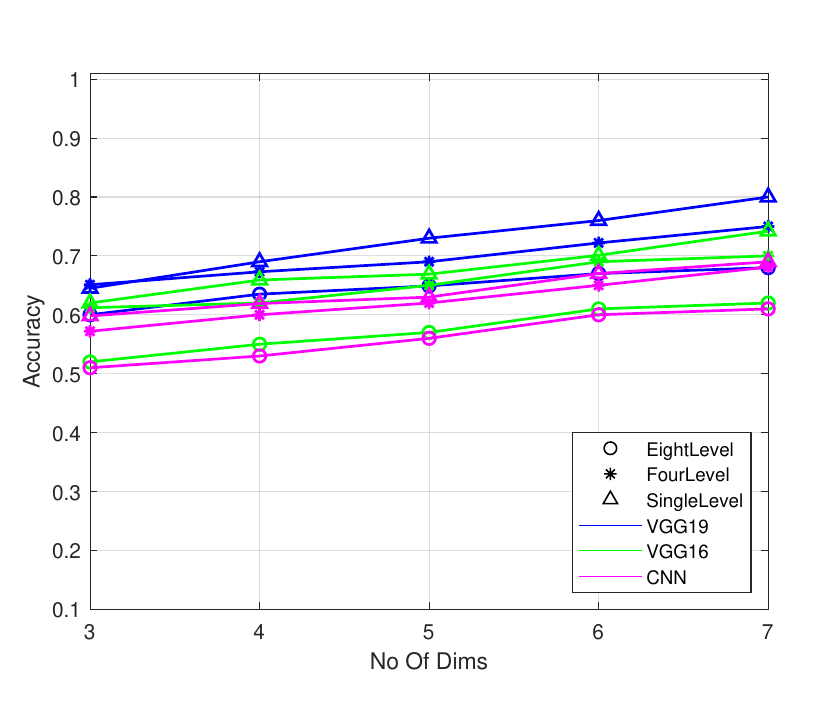}
		\captionsetup{justification=centering}
		\caption{CelebA}
		\label{fig:celeba_acc}
	\end{subfigure}
	\caption{Accuracy for Different Number of Reduced Dimensions }
	\label{fig:acc}
\end{figure*}

\begin{figure*}[ht!]
	\begin{subfigure}{.33\textwidth}
		\includegraphics[width=\linewidth, trim=3.8cm 8cm 4cm 8cm, clip=true]{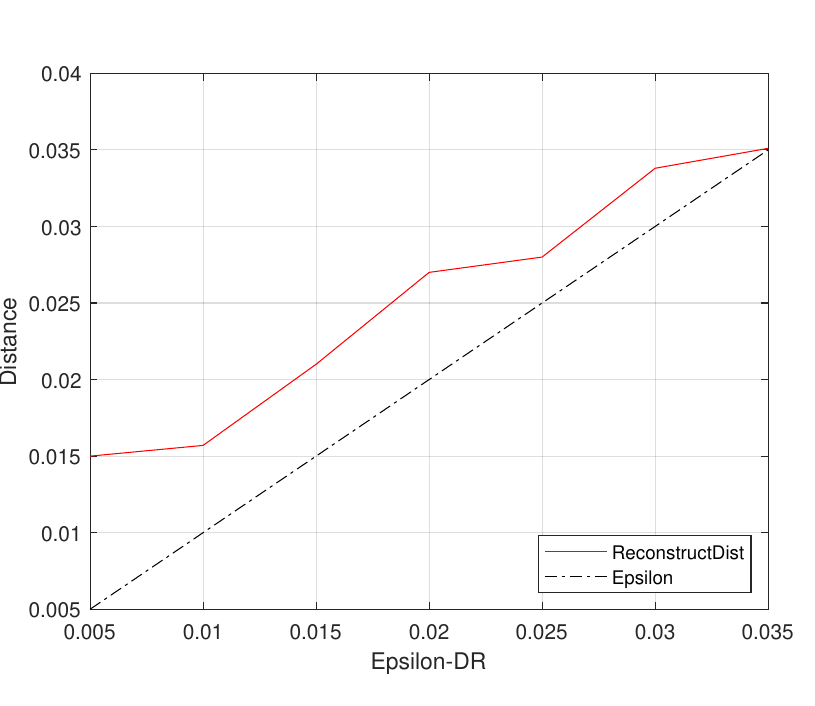}
		\captionsetup{justification=centering}
		\caption{ AT\&T}
		\label{fig:att_dist}
	\end{subfigure}
	\begin{subfigure}{.33\textwidth}
		\includegraphics[width=\linewidth, trim=3.8cm 8cm 4cm 8cm, clip=true]{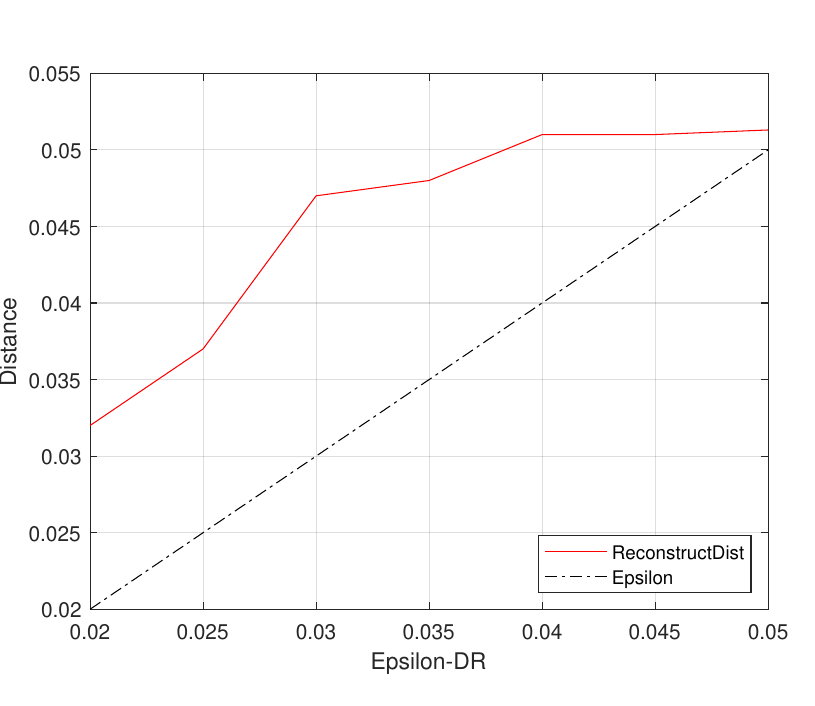}
		\captionsetup{justification=centering}
		\caption{Yale\_B}
		\label{fig:yale_dist}
	\end{subfigure}
	\begin{subfigure}{.33\textwidth}
		\includegraphics[width=\linewidth, trim=3.8cm 8cm 4cm 8cm, clip=true]{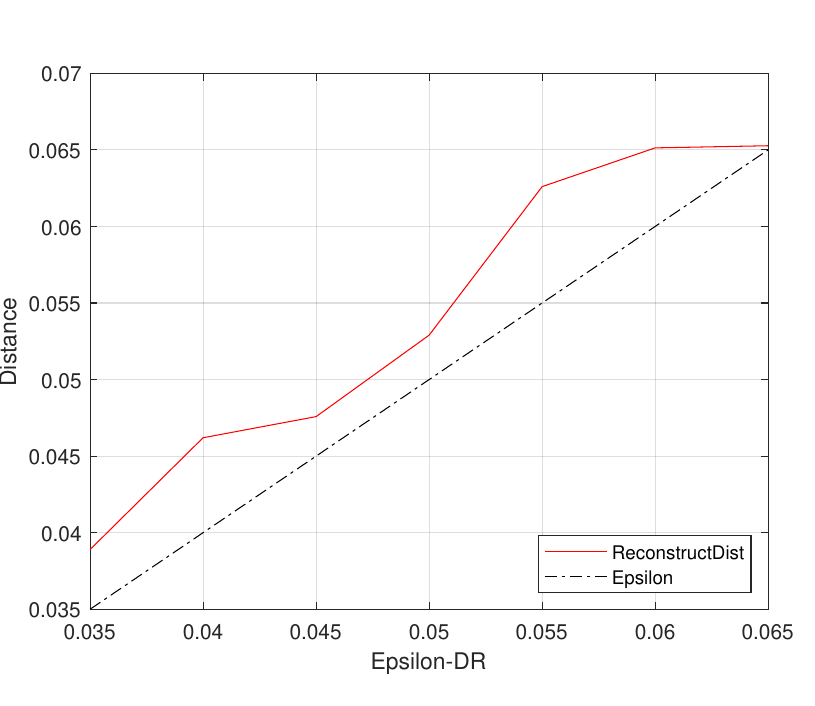}
		\captionsetup{justification=centering}
		\caption{CelebA}
		\label{fig:celeba_dist}
	\end{subfigure}
	\caption{Average Distance Measurement Result \{ 7 dimensions, Single-Level\}}
	\label{fig:distresult}
\end{figure*}

In this section, we demonstrate our experiments over three popular supervised face image datasets: \textit{the Extended Yale Face Database B} \cite{GeBeKr01}, \textit{AT}\&\textit{T} \cite{341300}, and \textit{CelebFaces Attributes Dataset (CelebA)} \cite{celeba}. To comprehensively evaluate our method performance, we also conduct experiments with different generator and re-constructor structures, different types of classifications (binary and multi-class classification), different numbers of reduced dimensions. The effectiveness of the method is then evaluated in terms of utility and privacy.   
\subsection{Experiment Setup}
\textit{The Extended Yale Face Database B} (YaleB) contains 2,470 grayscale images of 38 human subjects under different illumination conditions and their identity label. In this dataset, the image size is 168$\times$192 pixels. The AT\&T dataset has 400 face images of 40 subjects. For convenience, we resize each image of these two dataset to 64$\times$64 pixels. CelebA is a color facial image dataset containing 202,599 images of 10,177 subjects. 1,709 images of the first 80 subjects are used for our experiment. Each image is resized to 64$\times$64$\times$3 pixels. All pixel values are scaled to the range of [0,1]. We randomly select 10\% of each subject's images for validation and 15\% for testing dataset. 

The generator and re-constructor in Figure \ref{fig:eGAN} are implemented by three different structures. Specifically, we follow the architecture of recent powerful models VGG19, VGG16 \cite{vgg} and a basic convolutional network (CNN). We modify the models to adapt to our data size (64$\times$64). Discriminator and Classifier are built on fully connected neural network and convolutional network respectively. Leaky ReLU is used for activation function in hidden layers. We use linear activation function for generator's output layers and softmax activation functions for other components' output layers. Each component is trained in 5 local iterations ($n_r, n_g, n_d, n_c$), and the entire system is trained in 500 global iterations ($n$). The target distribution is drawn from Gaussian distribution (with the covariance value of 0.5 and the mean is the average of the training data). Table \ref{table:implementation} provides detail information of neural networks' structures and other implementation information. 
 
To evaluate the reliability, we test our framework with different levels of authentication corresponding to binary classification (single-level) and multi-class classification (multi-level). For the single-level authentication system, we consider half of the subjects in the dataset are valid to access company's resources while the rest are invalid. We randomly divide the dataset into two groups of subjects and labels their images to (1) or (0) depending on their access permission. For the cases of multi-level authentication system, we divide the subjects into four groups and eight groups. Therefore, the authentication server becomes four-class and eight-class classifier respectively. 
\subsection{Utility}
We use accuracy metric to evaluate the utility of dimension-reduced data. The testing dataset is tested with the classifier extracted from our framework. Different structures of Generator and re-constructor are applied including VGG19, VGG16, basic CNN on different privilege levels which correspond to multi-class classification. Figure \ref{fig:acc} illustrates the accuracies for different dimensions from three to seven over the three facial datasets. Overall, the accuracies improve when the number of dimension increases. The accuracies on the two gray image datasets (AT\&T and Yale\_B) reaches 90\% and higher when using VGG with only seven dimensions. This accuracy figure for Celeba is smaller, but it still reaches 80\%. In general, VGG19 structure performs better than using VGG16 and basic CNN in terms of utility due to the complexity (table \ref{table:implementation}) and adaptability to image datasets of VGG19. As the dimension number is reduced from 4,096 (64$\times$64) to 7, we can achieve a compression ratio of 585 yet achieve accuracy of 90\% for the two gray datasets and 80\% for the color dataset. This implies our method could gain a high compression ratio and maintain a high utility in terms of accuracy. During conducting experiments we also observe that the accuracy could be higher if we keep the original resolution of images. However, for convenience and reducing the complexity of our structure, we resize images to the size of 64$\times$64 pixels.    

\subsection{Privacy}
In this study, the Euclidean distance is used to measure the distance between original and reconstructed images: $dist(x,\hat{x}) = ||x-\hat{x}||^2$. Figure \ref{fig:distresult} illustrates the average distances between original images and reconstructed images on testing data with different $\epsilon$ constraints (other setting parameters: seven dimensions, single-level authentication, and VGG19 structure). The achieved distances (red lines) are larger than the hyper-parameter $\epsilon$ (black dotted lines) where $\epsilon$ is less than 0.035 for AT\&T, 0.052 for YaleB and 0.067 for CelebA. Thus, our framework can satisfy $\epsilon$-DR with $\epsilon$ of above values. Due to the fact that the re-constructor obtained some information (we consider the adversary can reach the model and the training data), we can only set the distance constraint $\epsilon$ within a certain range as shown in \ref{fig:distresult}. The intersection between the red line and the dotted black line points out the largest distance our framework can achieve. Since the mean of the target distribution is set to be the same as the mean of training dataset, reconstructed images will be close to the mean of training dataset which we believe it will enlarge the distance and expose less individual information. Thus, the range of epsilon can be estimated base on the expectation of the distance between testing samples and the mean of training data. In addition, the first section of Table \ref{table:visualization} demonstrates some samples and their corresponding reconstructions in single-level authentication and seven dimensions with different achieved accuracies and distances. The reconstructed images could be nearly identical, thus making it visually difficult to recognize the identity of an individual.      

\section{Comparison to GAP\cite{GAP}}
\label{AutoGAN_GAP}

In this section, we compare the proposed framework with GAP, which shares many similarities. At first, we attempt to visualize AutoGAN-DRP and GAP by highlighting their similarities and differences. Then, we exhibit our experiment results of the two methods on the same dataset. 

In terms of similarities, AutoGAN-DRP and GAP are utilizing minimax algorithms of Generative Adversarial Nets, applying the state-of-the-art convolution neural nets for image datasets, considering $l_2$ norm distance (i.e., distortion in GAP, privacy measurement in AutoGAN-DRP) between the original images and reconstructed images. Specifically, both GAP and AutoGAN-DRP consider the reconstruction distance between original and reconstructed images. In GAP this \textit{distortion} refers to the Euclidean between original and privatized images, and AutoGAN-DRP denotes the \textit{distance} as the Euclidean distance between original and reconstructed images. In this context, the distance and distortion refer to the same measurement and have the same meaning. To be consistent, we use the term \textit{distance} to present this measurement in the rest of this section.

However, there are also distinctions between GAP and AutoGAN-DRP. In GAP, the adversary aims to identify a private label (e.g., gender) which should be kept secret while AutoGAN-DRP aims to visually protect the owner's face images by enlarging the reconstruction distance. Thus, instead of considering a private label in loss function of the generator in GAP, AutoGAN-DRP is aimed at driving the reconstructed data into a target distribution using a discriminator.   
   
\begin{figure}
	\includegraphics[width=\linewidth,trim=2.7cm 0.5cm 2.5cm 0.5cm, clip=true]{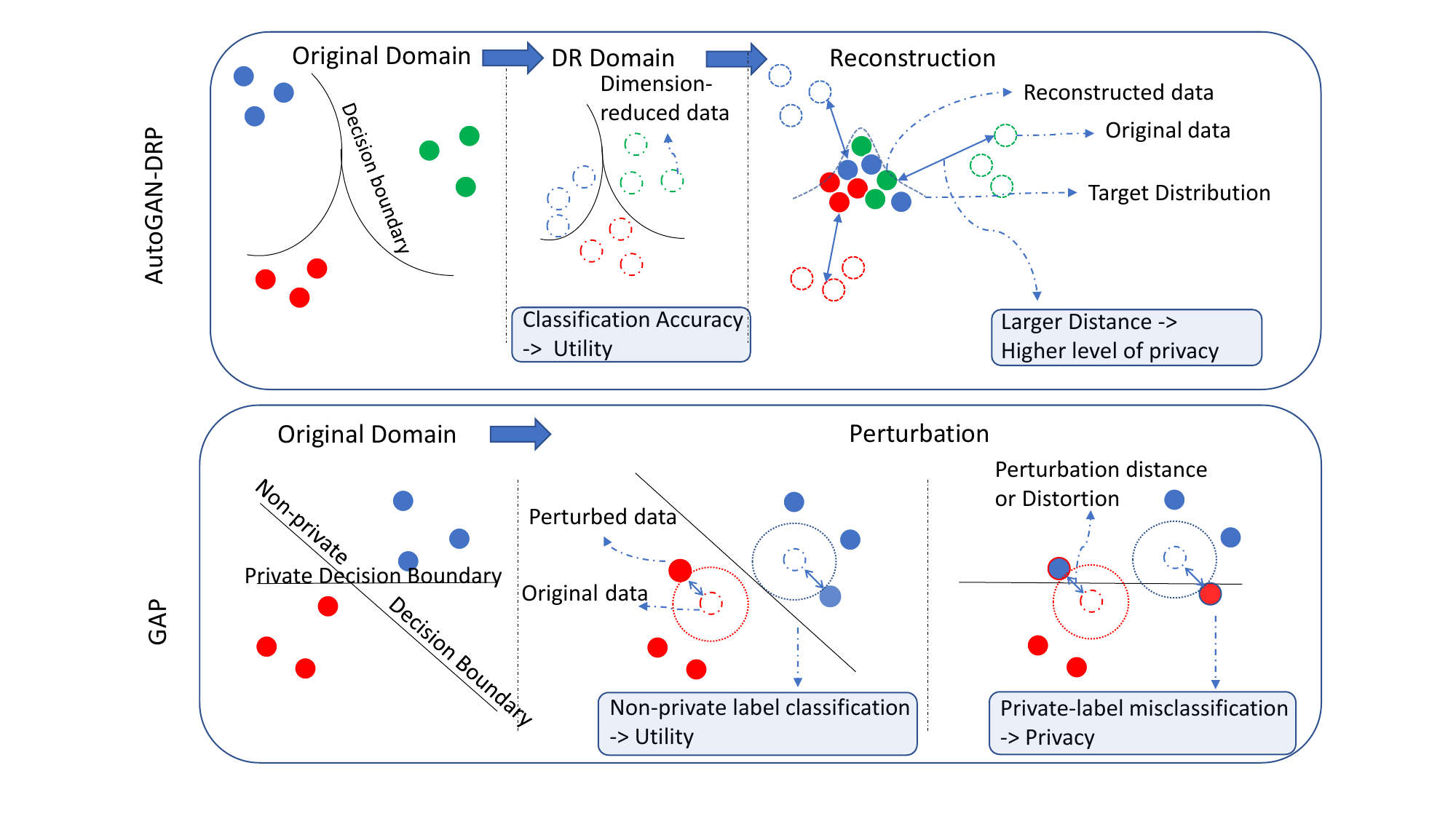}
	\captionsetup{justification=centering}
	\caption{AutoGAN-DRP Vs GAP Explanation}
	\label{fig:AGDRPPvsGAP}
\end{figure}
      
\begin{figure}[H]
	\includegraphics[width=\linewidth]{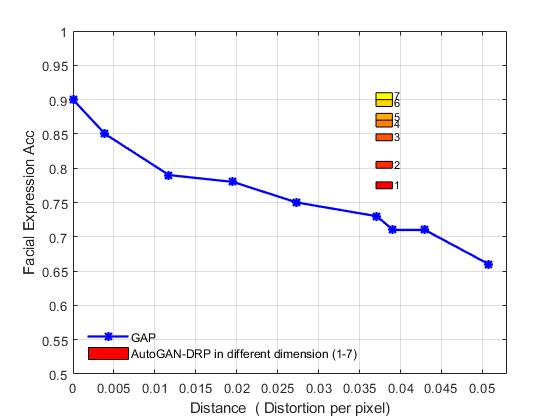}
	\captionsetup{justification=centering}
	\caption{GENKI Facial Expression Accuracy Vs Distance using GAP and AutoGAN-DRP}
	\label{fig:genki}
\end{figure}

Figure \ref{fig:AGDRPPvsGAP} illustrates the visualization of AutoGAN-DRP and GAP. In AutoGAN-DRP, privacy is assessed based on how well an adversary can reconstruct the original data and measured by the distance between original and reconstructed data. The dimension-reduced data is reconstructed using the state-of-the-art neural network (an Auto-encoder). The larger the distance is, the more privacy can be achieved. Further, if the reconstructed images are blurry, privacy can be preserved since it is hard to visually determine an individual identity. The data utility is quantified by the accuracy of the classification tasks over dimension-reduced data which captures the most significant data information. Meanwhile, GAP perturbs images with a certain distortion constraint to achieve privacy. It evaluates data utility by the classification accuracy of non-private label and assesses privacy by the classification accuracy of private label. Similar to AutoGAN-DRP, the high distortion is most likely to yield high level of privacy. In GAP, however, high distortion might dramatically reduce the classification accuracy of non-private label. This might be caused by the high correlation between private and non-private labels. This difference enables AutoGAN-DRP to preserve more utility than GAP at the same distortion level, as the experiment result (depicted in Figure \ref{fig:genki}) reveals. 

In the experiment, we reproduce a prototype of Transposed Convolutional Neural Nets Privatizer (TCNNP) in GAP using materials and source code provided by \cite{GAP}. We also modify our framework to make it as similar to TCNNP as possible. Specifically, a combination of two convolutional layers with ReLU activation function and two fully connected neural network layers are used for implementing the Generator similar to TCNNP. Our Classifier is constructed on two convolutional layers and two fully connected hidden layers similar to the Adversary in GAP. We also test our framework on GENKI, the same dataset with GAP. The utility is evaluated by the accuracy of facial expression classification (a binary classification). It should be noted that our framework have been shown to work on different datasets with multi-class classification, which is more challenging and comprehensive. Figure \ref{fig:genki} shows the accuracy results of GAP and AutoGAN-DRP for GENKI dataset. AutoGAN-DRP achieves distances ranging from 0.037 to 0.039 for different dimensions from one to seven. At the same range of distance (distortion per pixel), GAP achieves accuracy of only 72\% while AutoGAN-DRP gains accuracy rates starting from 77\% to 91\% for different number of dimensions. It becomes evident that our method can achieve higher accuracy than that of GAP at the same distortion level.

\section{Visual comparison to privacy preserving techniques using Differential Privacy (DP) \cite{Dwork2006} and Principle Component Analysis (PCA) \cite{PCA}}
\label{AutoGAN_DP_PCA}

In this section, we compare AutoGAN-DRP with other privacy preserving methods in terms of ability to visually identify client's identities. We choose the widely used tool for privacy preserving Differential Privacy (DP) \cite{Dwork2006} and another privacy preservation method utilizing dimensionality reduction technique (i.e., Principle Component Analysis \cite{PCA} ).
  
In these experiments, we implement AutoGAN-DRP following VGG19 structure for the Generator and Re-constructor, and other setting parameters (e.g., number of hidden layers, learning rate, optimization) are shown in Table \ref{table:implementation}. The images are reduced to seven dimensions for different values of $\epsilon$-DR to achieve different distances and accuracies. The datasets are grouped into two groups corresponding to a binary classifier. 

For implementing DP, we first generate a classifier on the authentication server by training the datasets with a VGG19 binary classifier (the structure of hidden layers is similar to our Generator in Table \ref{table:implementation}). The testing images are then perturbed using differential privacy method. Specifically, Laplace noise is added to the images with the sensitivity coefficient of 1 (it is computed by the maximum range value of each pixel [0,1]) and different DP epsilon parameters (this DP epsilon is different from our $\epsilon$-DR). The perturbed images are then sent to the authentication server and fed to the classifier. We visually compare the perturbed images of this method with AutoGAN. 

In addition, we follow instruction in FRiPAL \cite{zhuang2017fripal} in which the clients reduce image dimension using Principle Component Analysis (PCA) and send reduced features to the server. FRiPAL claims that by reducing image dimension, their method can be more resilient to reconstruction attacks. The experiments are conducted with different number of reduced dimension. The images are reconstructed using \textit{Moore–Penrose inverse} method with assumption that an adversary has assess to the model. The classification accuracy is evaluated using a classifier which has similar structure to AutoGAN's classifier. 

\begin{table*}
	\centering
	\includegraphics[width=0.9\linewidth, trim=1cm 3cm 1cm 3cm, clip=true]{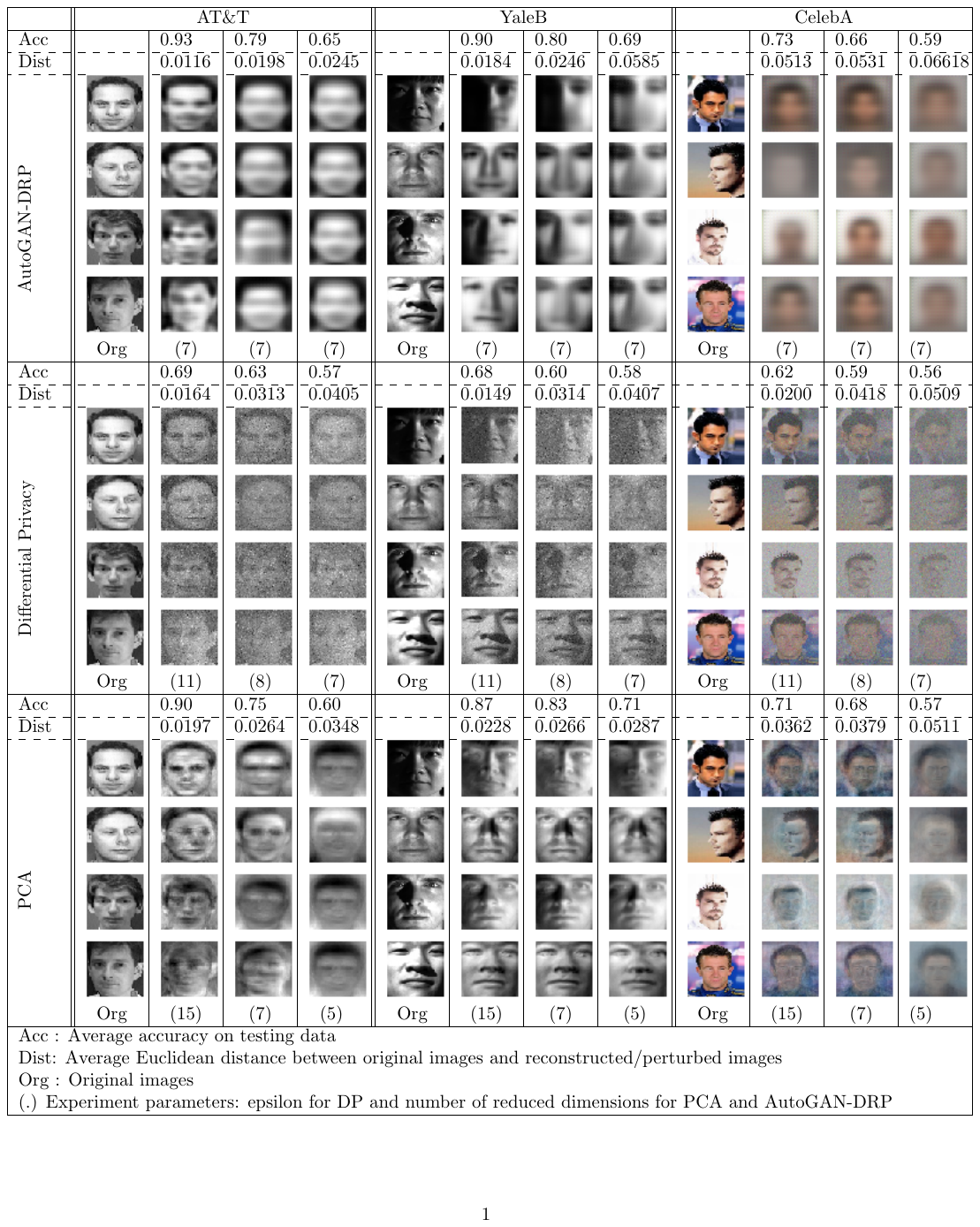}
	\caption{Sample visualization of AutoGAN, DP, PCA over three datasets}
	\label{table:visualization}
\end{table*}

Table \ref{table:visualization} shows image samples and results over the three datasets. Overall, AutoGAN-DRP is more resilient to reconstruction attacks compared to the other two techniques. For instance, at the accuracy of 79\% on AT\&T dataset, 80\% on YaleB, and 73\% on CelebA, we cannot distinguish entities from the others. For DP method, the accuracy decreases when the DP epsilon decreases (adding more noise), and the perturbed images become harder to recognize. However, at a low accuracy 57\%, we are still able to distinguish identities by human eyes. The reason is that DP noise does not focus on the important visual pixels. For PCA, the accuracy also goes down when the number of dimensions decreases and the distances increase. Since PCA transformation is linear and deterministic, the original information can be significantly reconstructed using the inverse transformation deriving from the model or training data. Thus, at the accuracy of 75\% on AT\&T, 71\% on YaleB, and 68\% on CelebA, we still can differentiate individuals. Overall, our proposed method shows the advantage in securing the data while retaining high data utility.

	\section{Conclusion}
	In this paper, we introduce a mathematical tool $\epsilon$-DR to evaluate privacy preserving mechanisms. We also propose a non-linear dimension reduction framework. This framework projects data onto lower dimension domain in which it prevents reconstruction attacks and preserves data utility. The dimension-reduced data can be used effectively for the machine learning tasks such as classification. In our future works, we plan to extend the framework to adapt with different types of data, such as time series and categorical data. We will apply different metrics to compute the distance other than $l_2$ norm and investigate the framework on several applications in security systems and data collaborative contributed systems.  
	
	\section*{Acknowledgment}
	This work is sponsored by DARPA Brandeis program under agreement number N66001-15-C-4068. The views, options, and/or findings contained in this article/presentation are those of the author/presenter and should not be interpreted as representing the official views or policies, either expressed or implied, of the Defense Advanced Research Projects Agency or the Department of Defense.

	\bibliography{MyCollection}
	
	\balance 
	\begin{wrapfigure}{l}{24mm} 		\includegraphics[width=1.05in,height=1.8in,clip,keepaspectratio]{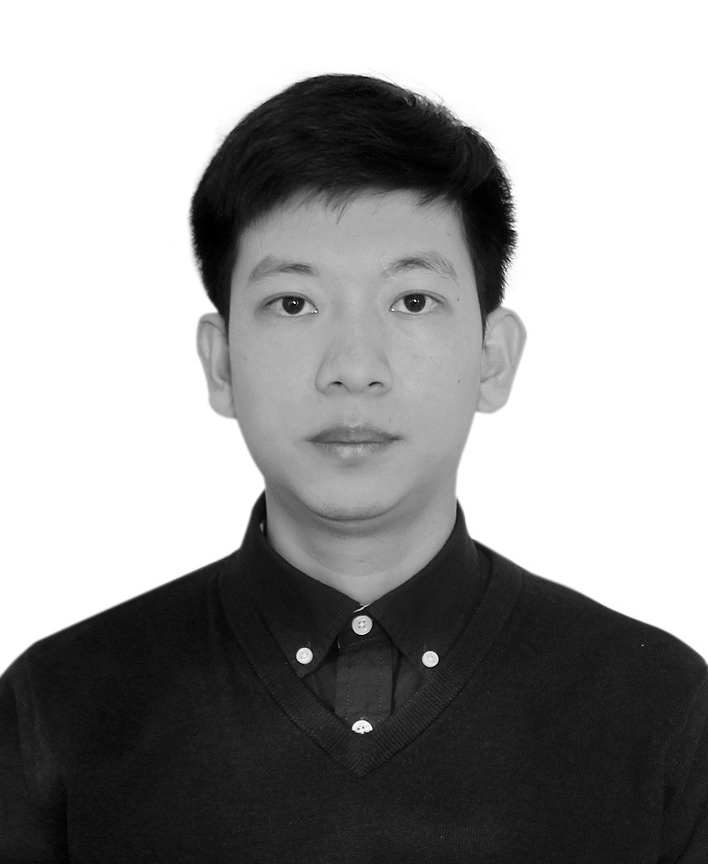}
	\end{wrapfigure}\par
	\textbf{Hung Nguyen} received the M.Sc. degree and he is currently pursuing his Ph.D. degree in Department of Electrical Engineering, University of South Florida, FL, USA. His current research interests include machine learning techniques, artificial intelligence, cyber security, and privacy enhancing technologies. He is a student member of IEEE.\par 
	
	\begin{wrapfigure}{l}{24mm} 
		\includegraphics[width=1.05in,height=1.8in,clip,keepaspectratio]{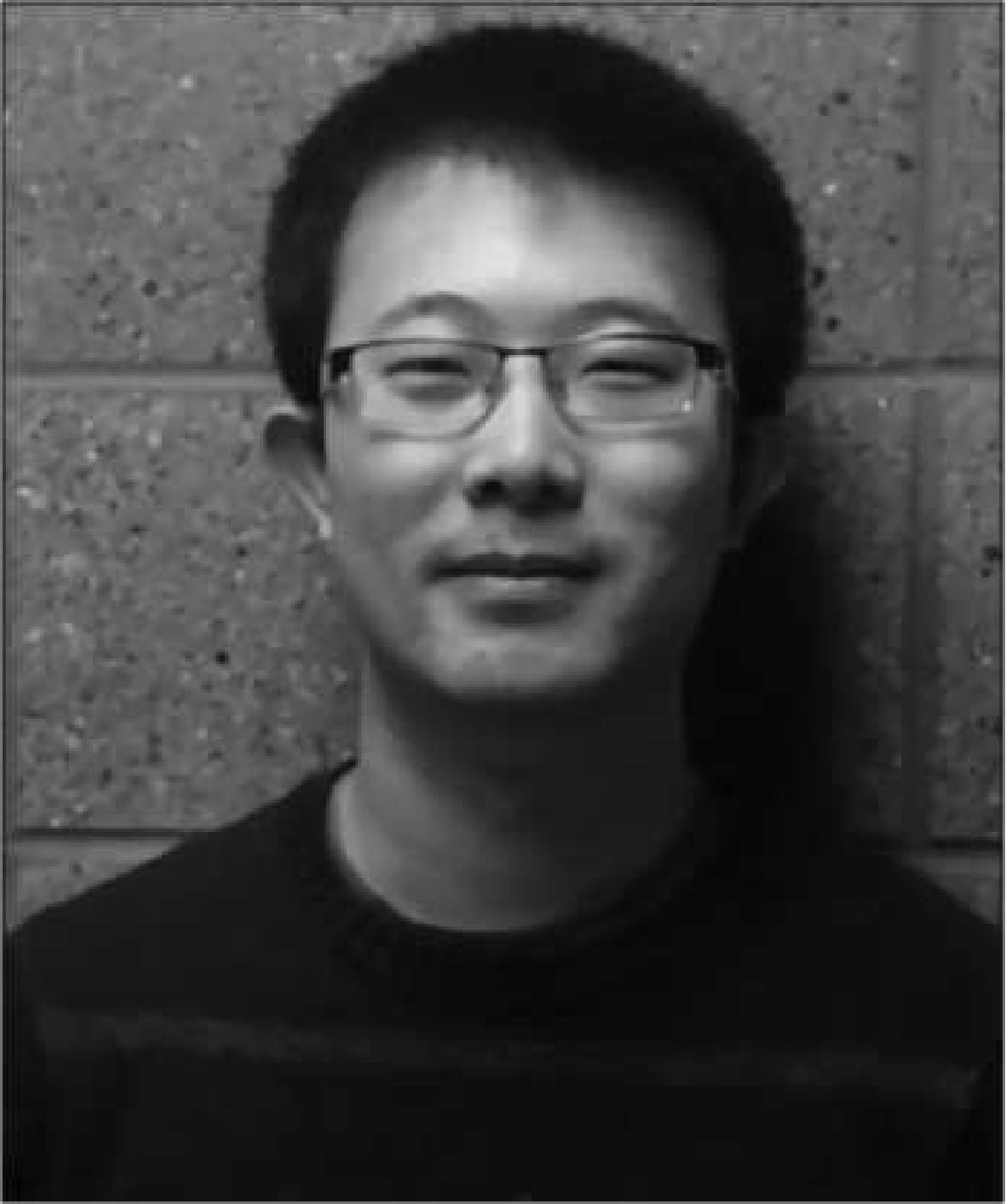}
	\end{wrapfigure}\par
	\textbf{Di Zhuang} received the B.E. degree in computer science and information security from Nankai University, China. He is currently pursuing his Ph.D. degree in electrical engineering with University of South Florida, Tampa. His research interests include cyber security, social network science, privacy enhancing technologies, machine learning and big data analytics. He is a student member of IEEE.\\ \par
	
	\begin{wrapfigure}{l}{24mm} 
		\includegraphics[width=1.05in,height=1.8in,clip,keepaspectratio]{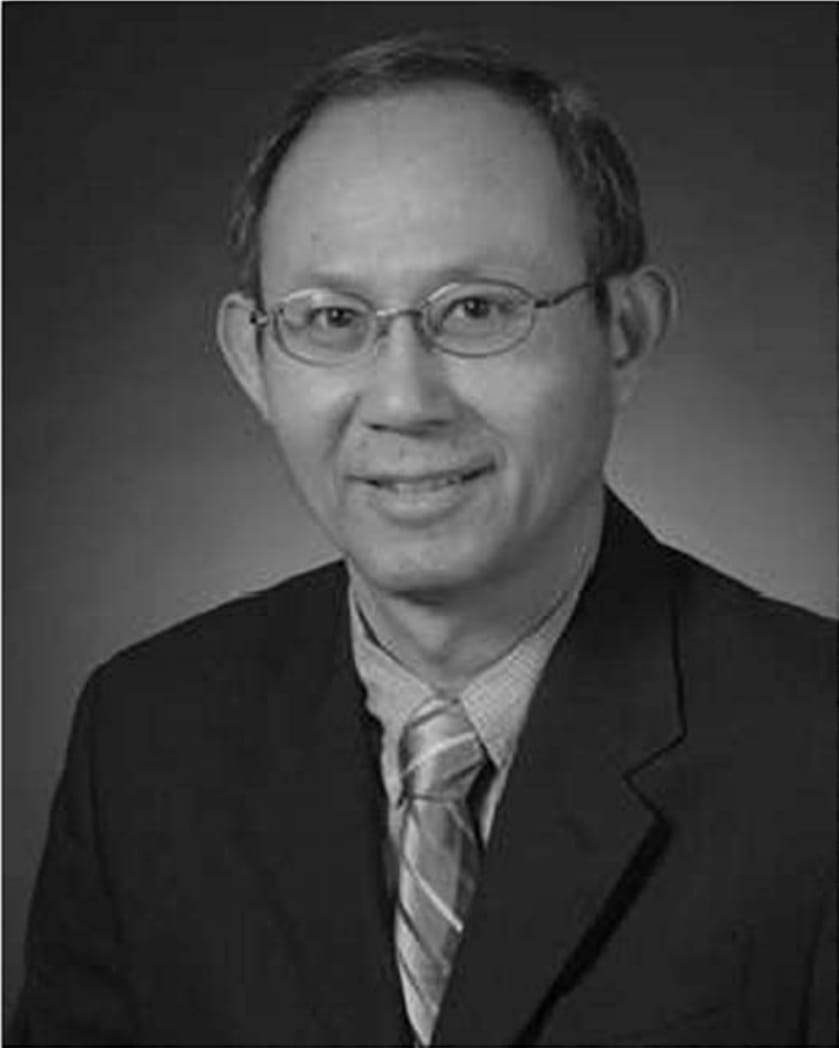}
	\end{wrapfigure}\par
	\textbf{J. Morris Chang} is a professor in the Department of Electrical Engineering at the University of South Florida. He received his Ph.D. degree from the North Carolina State University. He received the University Excellence in Teaching Award at Illinois Institute of Technology in 1999. His research interests include: cyber security, wireless networks, and energy efficient computer systems. In the last six years, his research projects on cyber security have been funded by DARPA. He is a handling editor of Journal of Microprocessors and Microsystems and an editor of IEEE IT Professional.\par

	\begin{wrapfigure}{l}{24mm} 
		\includegraphics[width=1.05in,height=1.8in,clip,keepaspectratio]{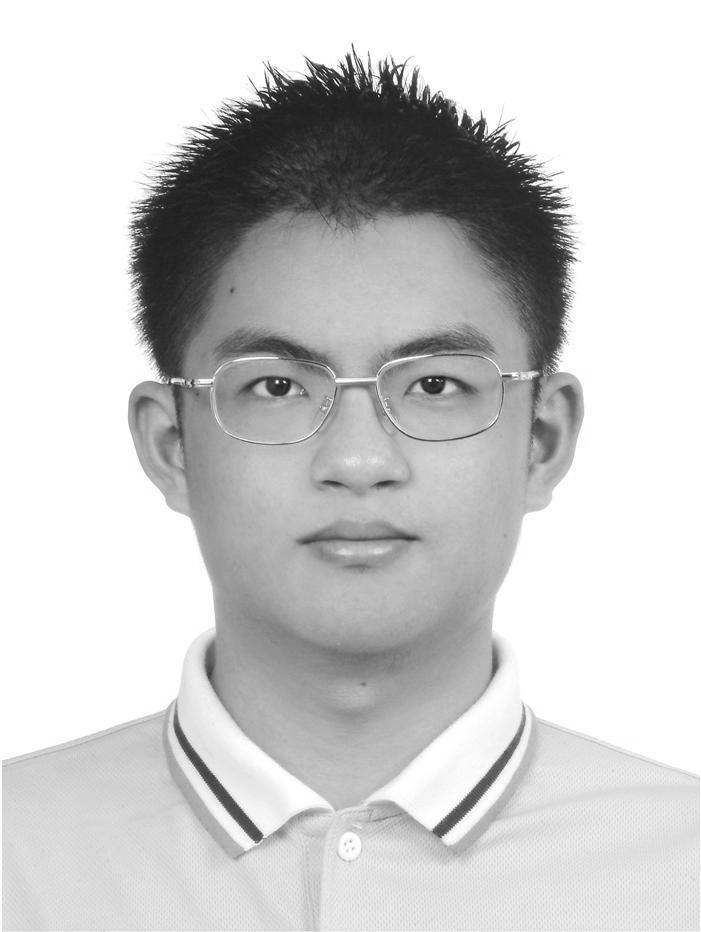}
	\end{wrapfigure}\par
	\textbf{PeiYuan Wu} is an assistant professor at National Taiwan University since 2017. He was born in Taipei, Taiwan, R.O.C., in 1987. He received the B.S.E. degree in electrical engineering from National Taiwan University in 2009, and the M.A. and Ph.D. degrees in electrical engineering from Princeton University in 2012 and 2015, respectively. He joined Taiwan Semiconductor Manufacturing Company from 2015 to 2017. He was a recipient of the Gordon Y.S. Wu Fellowship in 2010, Outstanding Teaching Assistant Award at Princeton University in 2012. His research interest lies in artificial intelligence, signal processing, estimation and prediction, and cyber-physical system modeling.\par
	
\end{document}